# Characterization of $WSe_2$ films using reflection Kikuchi diffraction in the scanning electron microscope and multivariate statistical analyses

Tianbi Zhang[1], Jakub Holzer[2], Tomáš Vystavěl[2], Miroslav Kolíbal[3,4], Estacio Paiva de Araujo[3], Chris Stephens[5], T. Ben Britton[1*]

1. Department of Materials Engineering, University of British Columbia, 309-6350 Stores Road, Vancouver BC, V6T 1Z4 Canada
2. Thermo Fisher Scientific, Vlastimila Pecha 1282/12, 627 00, Brno, Czech Republic
3. Central European Institute of Technology, Brno University of Technology, Purkynova 123, CZ-61200 Brno, Czech Republic
4. Institute of Physical Engineering, Brno University of Technology, Technická 2, 616 69 Brno, Czech Republic
5. Thermo Fisher Scientific, Materials Science, East Grinstead, United Kingdom

* Corresponding author: ben.britton@ubc.ca

# Abstract

The study of thin films and 2D materials, including transition metal dichalcogenides such as $WSe_2$ offers opportunities to leverage their properties in advanced sensors, quantum technologies, and device to optimize functional performance. In this work, we characterize thin $WSe_2$ samples with variable thicknesses using scanning electron microscope (SEM)-based techniques focused on analysis of backscattered electron signal and Kikuchi diffraction patterns. These data were collected via a pixelated electron-counting direct electron detector positioned below the pole piece primarily configured for reflection Kikuchi diffraction (RKD), and a similar detector placed in the more conventional electron backscatter diffraction geometry. In addition to conventional pattern analysis for orientation microscopy, multivariate statistical methods (MSA) based on principal component analysis were applied to analyze diffraction patterns and differentiate thickness variations and crystal orientations within the thin films through data clustering. These results were compared with atomic force microscopy to validate thickness measurements. Our findings indicate that RKD combined with MSA is highly effective for characterizing 2D materials, enabling simultaneous assessment of thickness and crystallographic orientation. Systematic acceleration voltage variations in RKD experiments and comparisons with EBSD data suggest that the thickness dependency arises from inelastic scattering of diffracted electrons, which affects pattern contrast in the thin film regime. Collection and analysis of patterns obtained from monolayer, bilayer and tri-layer of $WSe_2$ are also demonstrated. This work reinforces the utility of SEM-based techniques, such as RKD, as valuable tools for the materials characterization toolkit, particularly for thin films and 2D materials.

Transition metal dichalcogenides (TMDCs) such as $WSe_2$ are known for their unique electronic and optical properties and they have been identified as potential materials for e.g. thin film solar cells and flexible electronics [1–3]. These materials are often fabricated through exfoliation from bulk crystals or chemical vapour deposition onto a substrate. For these materials the thickness and grain structure of the film can be of interest for developing different types of electronic devices, as they can determine the properties of the film. For example, when the thickness of the film is reduced the band gap structure changes, and a monolayer $WSe_2$ film transitions to a direct band gap semiconductor [4,5], and transitions are also found in photoluminescence properties [6,7] and optical properties[8]. There is additional reporting of differences in the properties, including dielectric properties, of few-layer films and bulk materials [9]. Additionally, in these materials, grain size and its distribution can affect mechanical and electrical properties [10,11]. These facts highlight that characterization of film thickness and grain structure is an important task for both processing of raw materials, determining optimal engineering structures, and fabrication of optimized devices.

Grain structure of these films based on orientation information can be obtained from orientation microscopy techniques, such as scanning convergent beam electron diffraction (SCBED) and scanning precession electron diffraction (SPED) in the transmission electron microscope (TEM) [12–14], which generate spatially resolved orientation maps and offers additionally statistics on the size and shape of similar orientated domains. In terms of thickness, commonly used techniques include atomic force microscopy (AFM) [15], photoluminescence (PL) [16,17], Raman spectroscopy [17,18], electron energy loss spectrum (EELS) [19], 3D electron diffraction [20,21] and TEM imaging of the cross-section [22,23]. These techniques can be useful, but also have limitations in terms of mapping capability, film thickness range that can be investigated, or mapping speed.

Analyzing micrometer- or nanometer-scale features over large areas (μm to mm) is usually within the realm of optical or scanning electron microscope (SEM)-based techniques. Thickness of thin films can be qualitatively assessed from imaging based on intensity or contrast [24–26]. Orientation contrast can be revealed by Kikuchi diffraction techniques such as electron backscattered diffraction (EBSD), where the sample is tilted 70° towards a pixelated detector (Figure 1(a)) [27], and Kikuchi diffraction patterns are formed by backscattered electrons (BSE). Pattern analysis then allows reconstruction of the grain structure. However, the high sample tilt can result in image distortion and degradation of the physical spatial resolution [28]. To address these limitations, newer diffraction geometries can be used, such as transmission geometries, and the recently developed tilt-free geometry, namely "reflection Kikuchi diffraction (RKD)" [29], with a horizontal, pixelated detector placed around the pole piece of the SEM, and above a tilt-free sample (Figure 1(b)). These techniques have been more commonly used in characterization of nanomaterials [30–33].

Indexing of Kikuchi patterns typically only makes use diffraction features, especially the Kikuchi bands, to extract structural and orientation information, and thus thickness contrast, if any, is often overlooked in EBSD or RKD. In spite of this, it is well known that the overall BSE signal is affected by thickness of thin films [26], which can be observed in conventional SEM BSE imaging. RKD, which is geometrically similar to SEM BSE imaging, should then behave similarly and capture chemical, orientation, thickness contrast as a whole. From this perspective, another stream of pattern analysis can be conducted without conventional indexing, and instead using virtual diodes (or other motifs to form virtual apertures) on the patterns [34], statistical descriptors of the patterns [35], or statistical methods over the entire diffraction dataset, aiming to capture subtle contrast and features in the dataset in maps produced by these methods. For the latter, multivariate statistical analysis (MSA) can be of great importance for data clustering, because the pixelated RKD detector captures the BSE by a large number of pixels, where each pixel intensity is an independent variable sensitive to the contrast above, and the number of pixels is likely much larger than the number of classes in the dataset, i.e. regions with distinct orientation, structure or thickness. Evaluation of thickness from Kikuchi patterns additionally allows for studying the effect of thin film thickness on quality and contrast of backscattered Kikuchi patterns, which is much less studied than the thickness effect in their transmission geometry counterparts [36].

For Kikuchi diffraction datasets, especially EBSD datasets, the application of MSA-based method was first reported by Brewer et al. in 2008 using a principal component analysis (PCA)-based algorithm [37]. Wilkinson et al. [38] advanced this method in 2019 by incorporating the VARIMAX rotation to the PCA-identified components [39], so that each map point will be dominated by a component (a “characteristic diffraction pattern"). Both works were applied to segmenting grain structures (i.e. orientation-based) in metallic alloys. More recent improvements to this method include the works by McAuliffe et al. with considerations on the number of components selected, as well as incorporating energy-dispersive X-ray spectroscopy (EDS/EDX) data, for identification and classification of structurally similar second-phase particles, and comparisons with other MSA methods and machine learning methods were also performed [40,41].

In the present work, we extend the SEM-based RKD and EBSD analysis methods and combine it with a MSA method based on principal component analysis (PCA) and VARIMAX rotation, to classify RKD and EBSD datasets collected from multilayer and 1-3 layer $WSe_2$ thin films, and correlate them with conventional Kikuchi pattern analysis, and AFM- and photoluminescence (PL)-based film thickness measurements. In addition to typical orientation contrast, we will also demonstrate that the rich information contained in the RKD patterns can be used to cluster RKD datasets based on sample thickness and orientation using a MSA approach, which is a potential semi-quantitative technique to map the thickness and grain structure of crystalline thin films over a large area. Building upon the literature of electron scattering and diffraction in materials, we also comment on the origin of thickness, energy and geometry dependency of backscattered Kikuchi patterns in general, to gain further insights into the formation processes of Kikuchi patterns.

# Results

**Classification by MSA and validation.** We first study clustering of the RKD data from a thicker, multilayer film using MSA. Figure 1 shows the SEM image and inverse pole figure (IPF) map of the film. In these figures, orientation of the different regions from the flake can be observed, and there are variations in the SEM secondary electron signal that correspond to the grid and different regions across the sample. Two major flakes with a ~2° misorientation are identified from the IPF, which we refer to as “top flake” and “bottom flake” based on position.

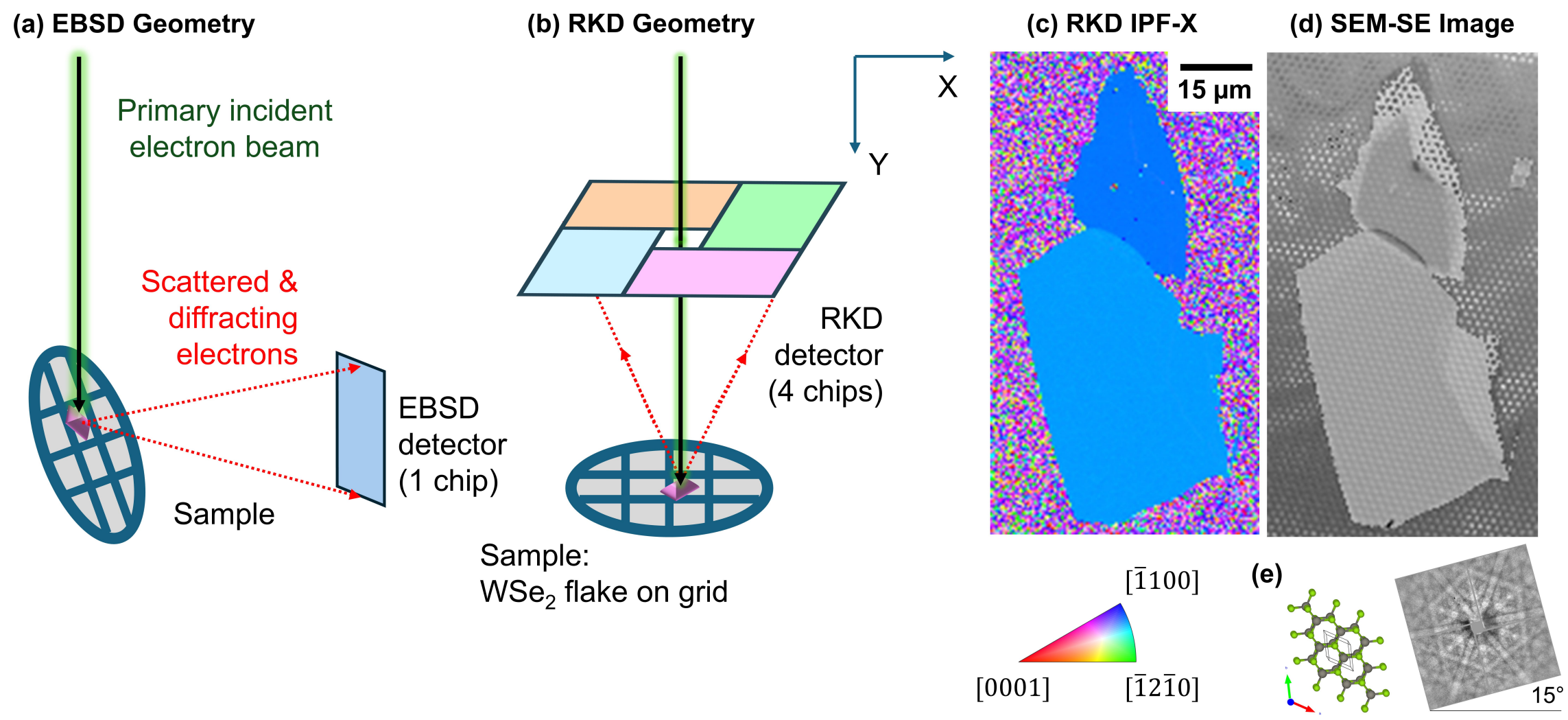

Figure 1. Schematic diagrams (not to scale) of the (a) EBSD and (b) RKD geometries, and (c) Inverse pole figure and (d) SEM image of the multilayer $WSe_2$ sample. (e) shows the atomic arrangement of $WSe_2$ and an example RKD pattern aligned to the orientation shown in (c). RKD data captured at 7 keV. $WSe_2$ flakes are supported on 200 nm holey $Si_3N_4$ membrane.

Figure 2 shows MSA analysis of the reference RKD collected at a primary electron energy of 7 keV. MSA performed on the full dataset with a variance tolerance of 0.01% (all components contribute a variance > 0.01% to the dataset) identified a total of 13 components, 3 are associated with the $WSe_2$ flakes (numbered F1-F3), and the rest are associated with non-diffracting features. While each of the characteristic patterns from the flake regions resemble real diffraction patterns, previous studies have suggested that these MSA-based components are statistical representations of the data, and as such do not necessarily have physical meaning. This can limit our analysis [37,38] and motivates us to build an approach to validate the MSA result and the nature of the characteristic patterns using a more physics-based analysis. First, the arithmetic summation of the physical RKD patterns collected from each of the three identified regions were calculated. Next, each of these summed patterns were compared to each experimental pattern in the dataset via image-based cross correlation. Figure 2(b) shows cross-correlation coefficient (XCC) maps for each summed pattern and the assignment map based on the highest ranked XCC for each point in the dataset. In general, the XCC-based assignment maps show similar results to the MSA-based maps, which assists in cross-validating the clustering based on the MSA approach.

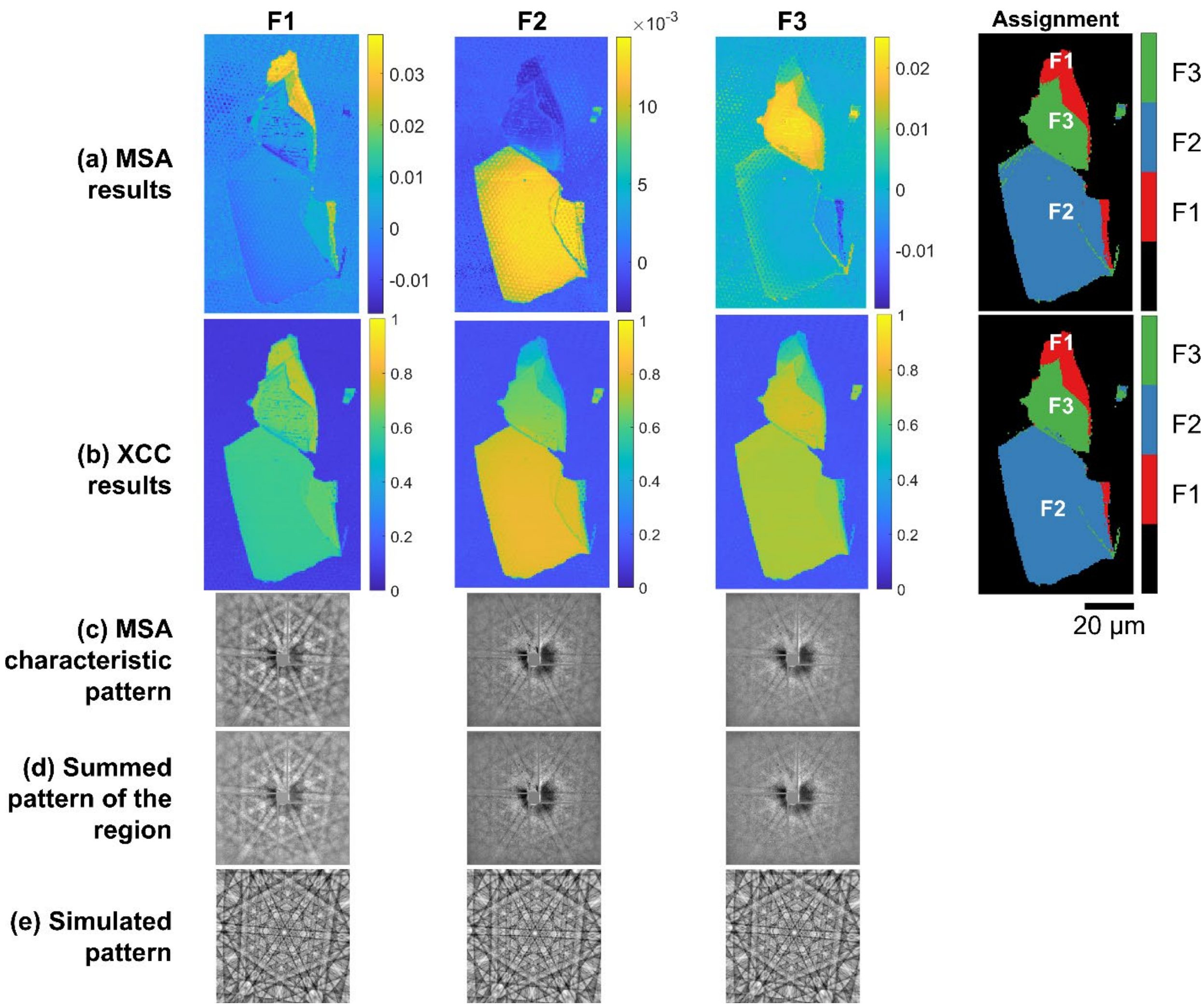

Figure 2. (a) Score maps of the components identified from MSA analysis and assignment map based on maximum score; (b) Normalized cross-correlation coefficient (XCC) maps compared to summed patterns of the MSA regions and assignment map. (c)-(e) shows the characteristic pattern, summed pattern of each region, and the corresponding simulated pattern based on indexing of (c). RKD data collected at 7 keV. $WSe_2$ flakes are supported on 200 nm holey $Si_3N_4$ membrane.

Film thickness from AFM measurements are tabulated in Table 1 and the number of layers is calculated assuming a layer thickness of 0.68 nm based upon prior HRTEM studies [22]. These results suggest that MSA clustering is based on orientation and thickness.

Table 1. Film thickness, average and standard deviation of median electron count (MEC) and "pattern quality" (PQ) statistics of the three identified regions. Standard deviation of MEC and PQ are reported in brackets.

| Region | AFM thickness (nm) | AFM thickness (#layers) | Median electron count | Pattern quality |
|---|---|---|---|---|
| F1 | 8-10 | 14-15 | 53.0 (17.5) | 69.9 (4.4) |
| F2 | 35-75 | 51-111 | 111.1 (7.3) | 75.7 (1.1) |
| F3 | 53-89 | 77-131 | 107.8 (19.4) | 73.5 (3.6) |

In the top flake, region F3 has a higher thickness of 53-89 nm and a higher average MEC of 107.8 and region F1 has a lower thickness of 8-10 nm and lower MEC of 53.0, as the yield of BSE increases with thickness for these thin film samples [26]. In the bottom flake, region F2 has a thickness of 35-75 nm and an average MEC of 111.1, and it is identified as a separate region from region F3 due to a small misorientation of around 2°. A thin strip to the right of region F2 was identified as region F1. Both MEC and PQ increase from region F1 to region F2 as shown by line profile analysis in Figure 3(a).

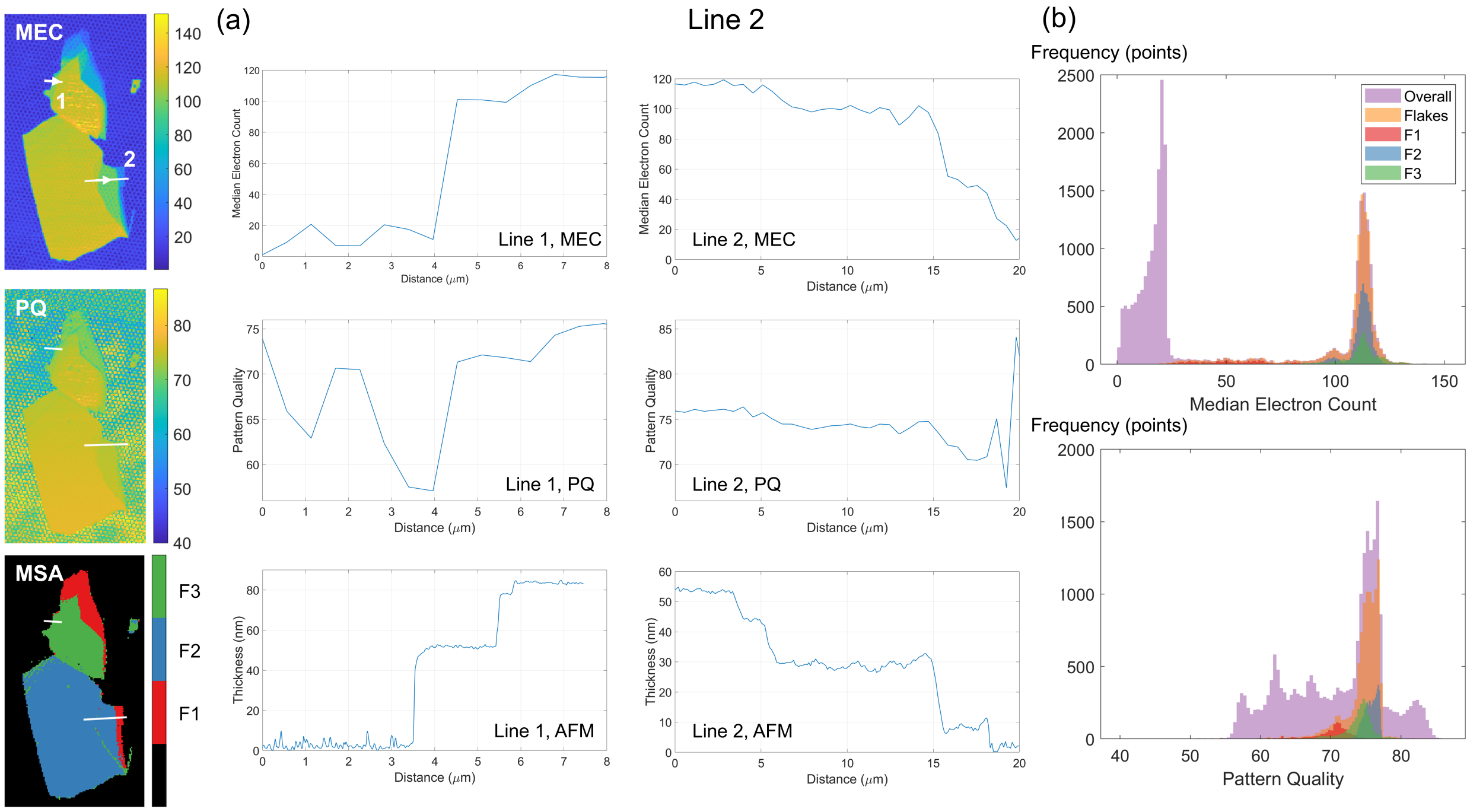

Figure 3. (a) Line profiles of normalized median electron count, pattern quality and thickness. Locations of the lines are indicated on MEC and PQ maps. (b) Distribution of median electron count and pattern quality over the entire flake region, and in each of the three regions identified by MSA, showing strongly overlapped peaks. RKD data collected at 7 keV. $WSe_2$ flakes are supported on 200 nm holey $Si_3N_4$ membrane.

Figure 3(b) shows the distribution of MEC and PQ of the entire dataset, of the flake regions and in each of the three flake regions. Even though differences in MEC and PQ are shown in the line profiles, the distribution of MEC and PQ of the entire dataset, of the flake regions and in each of the three flake regions overlap, especially MEC, which is also overlapped with that from the non-diffracting region. Thus, conventional classification methods such as multi-Otsu method based on MEC and/or PQ may become ineffective in both identifying the useful diffraction signal and clustering based on orientation and thickness contrast. This highlights the usefulness of the present MSA approach as the much larger number of observed variables overdetermines the classification problem. To examine the effect of diffraction geometry, MSA was performed on an EBSD dataset collected at 10 keV in a typical EBSD geometry (70° incidence angle, Figure 1(e)) from the same multilayer film. Using the same variance tolerance of 0.01%, only 5 components were identified from the full data set (c.f. 13 in RKD), and thickness contrast, especially on the top flake, is not resolved (more detailed MSA results can be found in the supplementary information).

**Towards monolayer and few-layer films.** Generally, it is believed that a sample must be thick enough for at least one quasi-elastic and incoherent scattering event to occur to give rise to Kikuchi patterns [42], and often at least few-layer structures are required for analysis by spot diffraction patterns from coherent scattering [20]. We demonstrate in Figure 4 that backscattered Kikuchi patterns from 1-3 $WSe_2$ layers can be captured, with thickness verified by photoluminescence (PL). In this case, the overall BSE signal shows additional contribution from the shadow of the TEM grid, and the RKD pattern is also influenced by whether the film is supported by the grid, or it is over a hole on the grid (i.e. not in contact with the grid).

Presence of the shadow in the patterns can be detrimental to line detection-based indexing, especially when the shadow contains regular line patterns (typical for many TEM grids and supporting substrates). On the software level, it is anticipated that more adaptive background correction routines can be useful to reduce the effect of the shadow to improve conventional indexing methods that are used during pattern collection, and patterns of sufficient quality can also be indexed by pattern matching. The pattern from the monolayer is evident but has low sharpness. We anticipate that pattern quality may be further improved with longer exposure times and benefit from careful studies of parameters such as detector energy threshold.

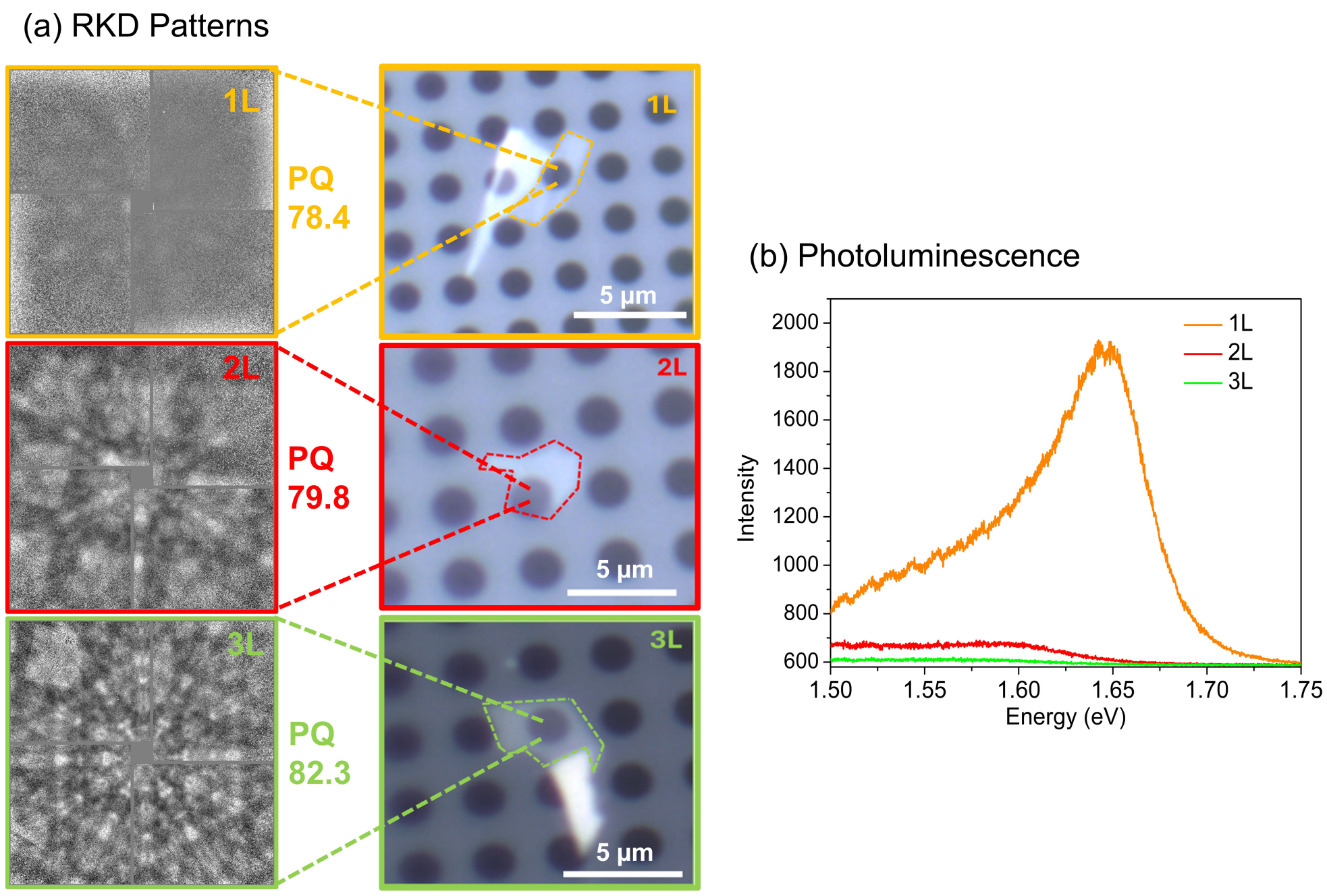

Figure 4. (a) Experimental RKD patterns at 8 keV electrons from 1-3 layers (1L-3L) of $WSe_2$ films, alongside optical micrographs. (b) PL spectrum of the monolayer exhibits a remarkably high intensity, which is a fingerprint of a direct band gap. Pronounced decrease in PL intensity is observed for the 2L and 3L samples due to the transition to an indirect band gap. $WSe_2$ flakes are supported on 20 nm holey carbon membrane.

# Discussions

It is remarkable that the RKD geometry enables new avenues of analysis of thin films, including measurement and understanding of the thickness of the film combined with full orientation information of the layer. This can be achieved using the calibrated electron counting information, cross correlation-based segmentation of the film, and statistical clustering methods. In this discussion, we outline likely mechanisms for the observed contrast in the RKD geometry, together with comparisons to the EBSD geometry, as well as use of the statistical methods to enhance our data processing pipeline.

**Mechanism of thickness dependency in Kikuchi patterns.** To briefly recap, Kikuchi patterns are (mostly) formed by an initial quasi-elastic, incoherent scattering of the incident electrons, followed by elastic and coherent scattering of electrons (Bragg scattering) by the lattice planes [43]. The thickness dependency of RKD pattern contrast that is observed in this work results from two competing regimes.

The mono- and few-layer samples represent the first regime, in which pattern quality and contrast are proportional to thickness. A higher sample thickness increases the tendency for quasi-elastic, incoherent scattering. In brief, “if the crystal is too thin, then (Kikuchi) diffraction cannot be strong” [44,45].

For the multilayer film with higher thickness the second regime prevails, where the number of quasi-elastic and incoherent scattering is no longer the limiting factor. Inelastic and incoherent scattering in the incoming path (prior to the quasi-elastic and incoherent step), and/or in the outgoing path (after the Bragg scattering step) are more likely. When such scattering occurs in the incoming path, the divergent electron source for Kikuchi diffraction will be distributed deeper in the sample, so we can assume that the distribution of the divergent source is directly related to sample thickness. Then after the final Kikuchi diffraction step, any inelastic and incoherent scattering in the outgoing path will reduce the contrast of Kikuchi bands [44,46], and a thicker sample with a longer outgoing path for electrons inside the sample will result in more of such scattering events and reduced pattern contrast. This is confirmed from the study of band intensity profiles of selected bands of the summed patterns of each region collected using 7 keV primary electron energy, shown in Figure 5. Band contrast is quantified using the difference between the highest and lowest band intensities within ±10° of the band opening angle and results are tabulated in Table 2. We note that the thinner region (F1) gives a consistently higher band contrast than the other two regions. This metric then provides an estimate of film thickness and is also sensitive to non-diffracting features, which will give a band contrast close to 0, and can be filtered out.

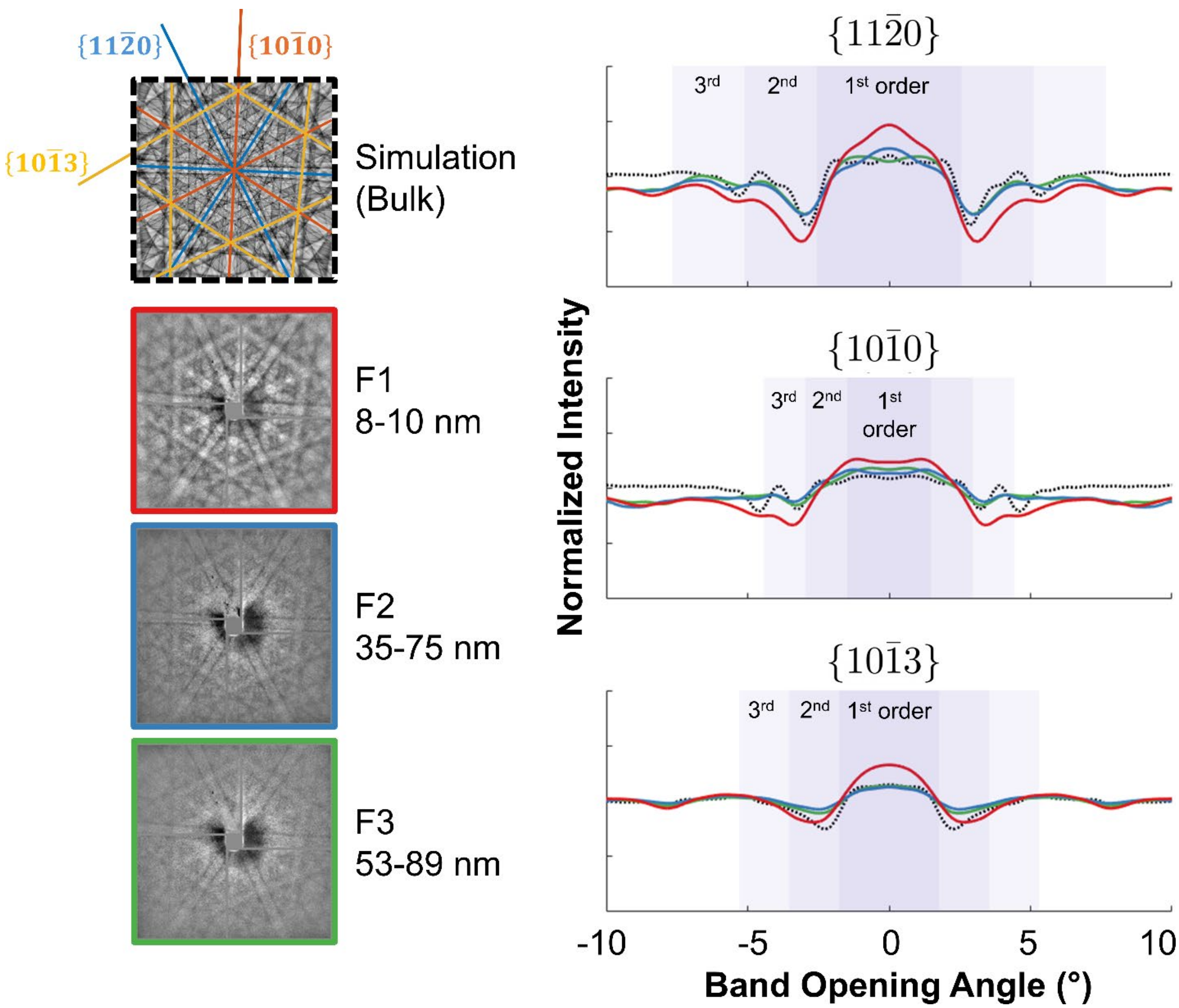

Figure 5. Normalized band intensity profiles of the $\{11\bar{2}0\}$, $\{10\bar{1}0\}$ and $\{10\bar{1}3\}$ bands on the summed experimental pattern of the three MSA regions of the multilayer sample and those of dynamically simulated patterns. Shading represents Bragg angle of first- and higher-order bands. RKD patterns and dynamical simulations were performed at 7 keV.

Table 2. Band contrast of the $\{11\bar{2}0\}$, $\{10\bar{1}0\}$ and $\{10\bar{1}3\}$ bands on the summed experimental pattern of the three MSA regions on the multilayer $WSe_2$ flake.

| Region | AFM thickness (nm) | Band Contrast | | |
|---|---|---|---|---|
| | | $\{11\bar{2}0\}$ | $\{10\bar{1}0\}$ | $\{10\bar{1}3\}$ |
| F1 | 8-10 | 1.054 | 0.596 | 0.518 |
| F2 | 35-75 | 0.521 | 0.336 | 0.245 |
| F3 | 53-89 | 0.600 | 0.291 | 0.201 |

As an outlook, a more formal and quantitative study could be performed with RKD on a sample with known and smooth thickness profile variation without influences from substrate and grid, akin to that by Brodu et al [36] , and for example the transition between the two regimes could be investigated as well.

**Thickness resolution.** The present MSA approach does not provide a direct measurement of film thickness or number of layers. Rather, since pattern contrast is affected by inelastic and incoherent scattering, the resolution of thickness clustering of our approach is likely related to the magnitude of the inelastic mean free path (IMFP) of electrons in $WSe_2$, which is estimated to be ~10 nm at 7 keV (based on the value at 80 keV in [47] and a square-root relationship [48]), and is far above the thickness of a monolayer. The effective resolution should be worse than the IMFP due to the probabilistic nature of scattering.

While this limits the thickness measurement capability of the present analysis, this analysis provides additional insights on diffraction physics of EBSD and RKD on thin film samples, which can complement the previous spectroscopy studies, and previous studies on the effect of diffraction geometry on pattern contrast [49]. MSA can also be performed on an individual grain (clustered by a common orientation) basis to isolate the thickness effect. With advanced dynamical pattern simulation capabilities, it may be possible to simulate thickness-dependent dynamical templates based on IMFP, absorption coefficients and other parameters to probe thickness contrast based on pattern matching methods.

Since the IMFP increases with electron energy (within the energy range of interest in this study), it is expected that in the second regime (multilayer films), pattern contrast should become more sensitive to sample thickness at lower primary electron beam energy until the film is thick enough to be considered a bulk material where thickness effects are diminished. To explore this hypothesis, we also performed the same MSA analysis on RKD datasets collected using 10, 15 and 20 keV primary electron energies (Figure 6). For the RKD datasets, thickness contrast in the top flake is consistently captured at every beam energy. However, compared to the 7 keV set, multiple components were identified on the thicker part of the bottom flake, and some can be attributed to whether the WSe2 flake is in contact with the $Si_3N_4$ membrane or over the holes. At 7 keV, effect of the grid on RKD pattern contrast and overall BSE yield is largely suppressed.

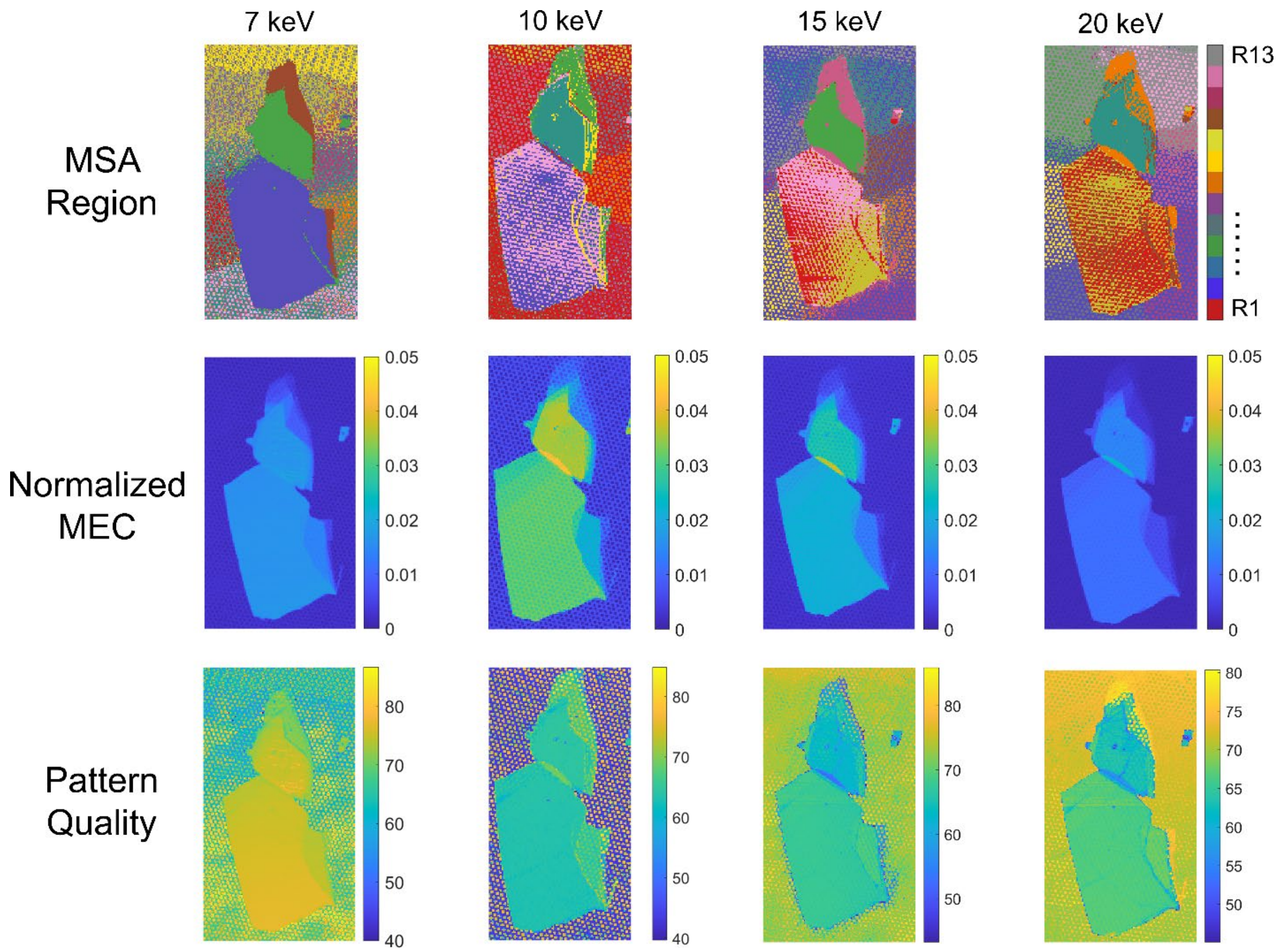

Figure 6. Maps of MSA component assignment, normalized MEC and pattern quality of the RKD datasets collected at 7, 10, 15 and 20 keV primary beam energies. Colors on the MSA maps represent different components. MEC data is normalized by the total incident electron dose (number of electrons). $WSe_2$ flakes are supported on 200 nm holey $Si_3N_4$ membrane.

While this result may promote the use of lower primary beam energies for similar RKD experiments, at lower electron energies, quality of the pattern can then be limited by low backscatter electron yield, and detector noise.

**Mechanism of geometry dependency in Kikuchi patterns**. The reduced sensitivity to thickness in the EBSD geometry can be attributed to the high sample tilt and detector placement. Due to the tilt, scattering in the sample occurs near surface. This impacts the scattering of the electrons as they enter and exit the sample, and the yield of backscattered electrons, meaning that for the EBSD geometry (as compared to the RKD geometry), the majority of inelastic scattering will occur in the incoming path, and the divergent electron sources that are required for Kikuchi diffraction will also be distributed closer to the sample surface. Critical to this understanding is the fact that band contrast and other Kikuchi diffraction related metrics, are affected by the scattering after the final scattering event [44], even though the number of scattered electrons and their energy will be affected by the energy spectrum of the incident scattered electron beam. By contrast, in the RKD geometry, divergent electron sources on average are distributed further from the surface, and inelastic scattering becomes more evenly distributed in incoming and outgoing directions, so their effects on band contrast are more prominent. This is why there is a greater sensitivity to film thickness for the RKD method.

Furthermore, while some EBSD patterns may still be affected by the electron scattering processes related to sample thickness, this is not readily observed in the MSA based clustering approach as patterns from different thickness regions on the multilayer sample are statistically more similar. This further indicates the EBSD pattern contrast shows a reduced sensitivity to thickness in comparison to RKD.

It should be emphasized that for orientation determination and grain structure mapping, both RKD and EBSD can be applicable, so long as an appropriate pattern analysis method is used. [49]

**Other experimental considerations.**

Beam current: The beam currents have been selected for these experiments to provide high-quality wide-angle diffraction patterns. Kikuchi diffraction, especially in the RKD geometry for electron transparent samples, is not efficient, because (a) many of the incident electrons are transmitted with little or no interaction with the sample (i.e. in the direct beam for a TEM experiment and contributing to the zero loss EELs peak), and (b) the efficiency of scattering to wide angles and the generation of Kikuchi bands is lower. These factors motivate longer exposure times and/or higher probe currents to resolve the bands well [49].

Voltage: The beam energies were selected mainly to overcome the background noise of the detector, determined by its minimum energy threshold (3-5 keV). Electrons with energy lower than the threshold will not be detected, while electrons with energies slightly above the minimum threshold will not be detected efficiently due to charge sharing effects. In addition, the effect of beam energy on electron probe size can also be taken into consideration for improving the physical spatial resolution [50,51].

Contamination: Most SEMs operate at moderately high vacuum and sample contamination may be an issue, typically due to the formation of a carbon rich layer on the surface of the sample. Contamination can be mitigated by general cleanliness when handling the sample and microscope (i.e. use of gloves and clean instrumentation), and the use of sample baking or plasma cleaner of the SEM chamber. The SEM instrument used in this study is equipped with a plasma cleaner to keep the chamber clean, but plasma cleaning was not used to clean the samples further to avoid risk of sample modification by the plasma. The contamination film can affect electron scattering, especially for fewlayer samples, but being amorphous in nature the contamination film should not affect diffraction information from the crystalline sample.

Shadow effect and substrate selection: Based on the diffraction geometry and paths of electrons, the shadow in the fewlayer sample patterns is likely the shadow of the Cu TEM grid, formed by transmitted electrons which are scattered inside the SEM chamber and back to the RKD detector. Similarly, in the multilayer sample, a shadow of the windowed substrate can also form, but with the presence of a stronger Kikuchi diffraction signal from the multilayer sample, the shadow is not readily observed (Figure S2). Note that the windowed substrate does not form a “criss-cross” shadow, and thus similar windowed substrates can be recommended to reduce the shadow effect for the fewlayer samples. To suppress the shadow, in theory a Faraday cup below the sample can be used to collect transmitted electrons, which requires modification of the SEM chamber; or a thicker and uniform substrate can be used to suppress transmitted electrons. Although for the latter option, the substrate will contribute even more to the diffraction signal than an electron transparent membrane. In this case, the energy threshold of the detector may be adjusted to reduce this effect. Meanwhile, the sample will not be suitable for TEM and related modalities, where a transmitted signal is to be observed.

**Outlook of the RKD+MSA approach.** In this work, the RKD+MSA approach allowed clustering of diffraction data based on the existence (or absence) of diffraction signal, orientation and thickness of thin film sample. Additionally, the approach has also helped in identifying the thickness dependency of Kikuchi pattern contrast in thin films which promotes and enables the use of SEM-based methodologies for rapid sample scanning and screening, alongside other characterization tools. As an example presented in this work, a region with common orientation identified by MSA, indexing the diffraction patterns from this region can be simplified to indexing the sum pattern of the region, along with any refinements required [38]. This has the advantage of amplifying the signal-to-noise ratio of the dataset, reducing the computation cost for pattern indexing, and also obtaining an effectively higher quality, “high dose” characteristic diffraction pattern using lower electron doses per experimental pattern and per unit area. This can lead to reduction of sample contamination and potential beam damage.

**Alternative clustering methods.** As was stated in the introduction, the present MSA routine provides the opportunity to use the PCA as a first decomposition, followed by rotation of a sub-set of the resolved components to achieve near uniform variance. This is useful as a first approach to cluster the data when the expected signals (i.e. components) are physically similar, even though the resulting components may not represent realistic features on the sample or in the diffraction data. More recent works have looked into non-linear clustering methods such as non-negative matrix factorization (NMF), which will always produce non-negative scores and has been shown to aid the interpretation of spectral data as well as improved clustering [41,52]. Machine learning approaches such as autoencoder have also been proposed [41,53]. In short, applications of decomposition methods in general to 4D diffraction datasets are still limited in scope at present and it is anticipated that, especially for thin films where the amount of signal generated is extremely small, it will be valuable to explore more systematic studies on the effect of each of the clustering methods (including variants of methods to improve robustness and interpretability of results) as well as data normalization methods. It is also worth noting that unlike 1D EDS spectra, Kikuchi patterns are 2D projections from a reference spherical pattern [54], and this projection pattern will also be sensitive to the position of the projection centre (and this is especially important when the map is larger than a detector pixel size (which is not the case for the present work, the detector pixels are 55 μm in size and the maps are all smaller than this). At present, the effect of large pattern centre shift on accuracy of any of the clustering methods mentioned above has yet to be assessed. In addition, in the current MSA approach, the user still decides on the number of classes to be extracted either explicitly or implicitly (e.g. using the variance tolerance). Individual classes identified by MSA may not necessarily result in meaningful clustering of the data, and appropriate selection of the number of classes may require trial and error and *a priori* knowledge on the sample.

# Conclusions

We demonstrated that the RKD technique can be used as a more advanced BSE imaging mode to reveal orientation information from conventional Kikuchi pattern analysis. Thickness contrast of $WSe_2$ thin films can also be revealed through a multivariate statistical analysis derived from the conventional PCA approach. Thickness contrast in Kikuchi patterns is caused by inelastic and incoherent scattering as diffracted electrons escape the sample, and higher thickness causes more energy loss and pattern contrast decrease. Thickness clustering qualitatively agrees with AFM measurements. The resolution of thickness classification based on the MSA approach is considered in the order of the inelastic mean free path of electrons. Using this approach, we demonstrated that with the help of an amorphous film, backscattered Kikuchi patterns can be captured from mono- and few-layer $WSe_2$ films.

## Methodology

**Materials.** $WSe_2$ flakes were prepared by the deterministic polydimethylsiloxane (PDMS) dry transfer method [55]. First, $WSe_2$ (purchased from 2D Semiconductors) was transferred onto PDMS via mechanical exfoliation using adhesive tape. Subsequently, the PDMS was mounted onto a glass slide connected to a micromanipulator (MPS 150). While monitoring the process through an optical microscope, the $WSe_2$ flakes were aligned onto the supporting membrane of a TEM grid (or a windowed substrate) and brought into contact. To release the $WSe_2$ flakes from the PDMS, the grid-PDMS sandwich was heated to 60 °C for approximately 3 minutes, after which the PDMS was slowly detached, leaving the flakes positioned at the desired location on the grid or substrate. The multilayer sample was deposited onto a 200 nm thick holey $Si_3N_4$ membrane on a 3 mm diameter windowed Si substrate (Ted Pella Inc, 21581-10). The few layer samples were deposited onto a 20 nm thick holey carbon membrane on a Cu TEM grid (Ted Pella, 658-200-Cu). No cleaning was performed on the samples for RKD and EBSD.

**Electron microscopy.** RKD and EBSD were performed in a Thermo Fisher Scientific Apreo II SEM at 7-20 keV primary electron energy, 6.4-12.8 nA beam current with pattern exposure times of 10-30 ms per pattern (the individual details of each experiment are provided in Supplementary Information). The TEM grids were mounted onto a bar holder (Figure S1(a)) using a spring to clamp the TEM grid. The holder is then mounted onto the sample stage of the SEM.

In the RKD geometry, no tilt was applied to the sample, so the surface of the sample is perpendicular to the electron beam, and Kikuchi patterns are collected by detector above the sample (Figure S1(b)). The detector used for these RKD experiment consists of a single crystal of Si with a central hole for the electron beam to pass through and four Timepix readouts tilted around it.

The RKD detector is inserted into the SEM chamber at an angle of 15° from the X-direction of the SEM scan grid (i.e. about the Z-sample axis, as reflected in Figure 1). This rotation is accounted for in the pattern indexing process. Raw RKD patterns are obtained by tiling the signals from the four detector chips to a single 549x549 array (with empty pixels corresponding physically to the gaps between the chips and the central hole region). Processed RKD patterns are obtained based on flat-fielding the central 479x479 region using the average “pattern” collected over a number of random positions in the field of view of the SEM.

For EBSD experiments, a TruePix direct electron detector was used which is based on a single Timepix chip with a Si sensor. The bar holder was mounted onto the SEM stage differently from the RKD geometry, and the SEM stage is tilted so that the net tilt of the sample towards the detector is 70°. Sensor of the EBSD detector is vertical (i.e. the detector plane contains the sample tilt axis and the axis of the primary electron beam, see Figure S1(c)). The TruePix detector collects patterns with 256x256 pixels. The background correction method for EBSD patterns is the same as in RKD. An energy threshold is applied to the detector to filter out low energy electrons in both RKD and EBSD experiments.

RKD and EBSD data collection and initial analysis were performed in the Thermo Fisher EBSD software (xTalView). This software also calculates the median electron count (MEC) obtained from the raw diffraction pattern as the median of the (calibrated) electron count registered among the pixels, and “pattern quality (PQ)”. PQ is calculated as a normalized signal-to-noise ratio. Specifically,

$$PQ = 100\left(1 - \frac{\frac{\sum_{i,j} FFT(I)\ M_O}{\sum_{i,j}\ M_O}}{\frac{\sum_{i,j} FFT(I)\ M_I}{\sum_{i,j}\ M_I}}\right)$$

Where $I$ is the processed pattern, $FFT(I)$ is the 2D fast Fourier transform power spectrum of $I$, $i,j$ is the position of a pixel on the processed pattern, $M_O$ is the “outer mask” which masks out the outer region of the pattern, and $M_i$ is the “inner mask” which masks out the central region of the 2D FFT power spectrum.

**Data processing and analysis.** Pattern data was analysed further within MATLAB using MTEX (version 5.11.1) [56] and AstroEBSD packages [57]. Within AstroEBSD, the MSA routine was conducted using the code developed by McAuliffe et al. [40] for EBSD datasets, which is based on the PCA + VARIMAX approach by Wilkinson et al. [38]. For MSA, RKD patterns were binned by a factor of two (from 479x479 to 240x240) and the EBSD patterns not binned. Next, they were vectorized and normalized to zero mean and unity standard deviation. In this manuscript, we refer to “component” as the characteristic signal identified from MSA, “characteristic pattern” as an identified characteristic component reshaped into a Kikuchi pattern-like image, and “region” as locations on the sample classified as a particular component based on the highest score of the components as compared to all other potential components. All data processing and analyses were performed in MATLAB.

Indexing of individual patterns were performed using the refined template matching method in AstroEBSD [58]. Dynamical templates of $WSe_2$ patterns were simulated based on the methodology described in [59] assuming a bulk sample. Band profiles were calculated based on pattern indexing, reprojection and the spherical harmonics approach [60]. Patterns are normalized to zero mean, unity variance, and a bandwidth of 384 was used for the spherical harmonic approximation. All symmetry equivalent bands are considered.

**Validation of film thickness.** A combination of photoluminescence (PL), and atomic force microscopy (AFM) was used to determine the number of layers for each transferred flake. In PL measurements (633 nm illuminating light), the PL intensity, peak shape and peak position were evaluated to discriminate flakes containing up to three layers [17]. PL measurements were performed using a WITec Alpha 300R system at room temperature. AFM was used to analyze thicker flakes (>3 layers); the measured flake height was divided by the interlayer distance of bulk $WSe_2$ (0.68 nm) to estimate the number of layers. AFM was performed with a Nanoscan VLS-80 instrument using a tapping mode with a Scanasyst-tip. Analysis was performed in Gwyddion v2.68.

# CRediT Authorship Contribution Statement

T. Zhang: investigation, methodology, data curation, formal analysis, visualization, validation, writing – original draft.

M. Kolíbal: investigation, methodology, data curation, formal analysis, writing – review & editing.

E. P. de Araujo: investigation, methodology, data curation, formal analysis, writing – review & editing.

Jakub Holzer: resources, investigation, methodology, data curation, formal analysis, visualization, validation, writing – review & editing.

Tomáš Vystavěl: project administration, resources, writing – review & editing

Chris Stephens: project administration, resources, writing – review & editing

T. B. Britton: conceptualization, funding acquisition, project administration, resources, supervision, visualization, writing – review & editing.

# Data Availability Statement

RKD and EBSD datasets are available on Zenodo at https://doi.org/10.5281/zenodo.15741760. MATLAB scripts for data processing and plotting are available in the AstroEBSD repository (https://github.com/ExpMicroMech/AstroEBSD).

# Conflict of Interest

J. Holzer, T. Vystavěl and C. Stephens are employees of Thermo Fisher Scientific, who make and distribute the Apreo SEM, RKD detector and TruePix EBSD detector used in this work.

# Acknowledgements

T.Z. and T. B. B. acknowledge the following funding support: Natural Sciences and Engineering Research Council of Canada (NSERC) [Discovery grant: RGPIN-2022-04762, ‘Advances in Data Driven Quantitative Materials Characterization’], Canada Foundation for Innovation and British Columbia Knowledge Fund (BCKDF) via the John

R. Evans Leaders Fund (CFI-JELF) [#43737, 3D-MARVIN]; CzechNanoLab project LM2023051 funded by the Ministry of Education, Youth and Sports of Czech Republic is gratefully acknowledged for the financial support of the measurements at Central European Institute of Technology (CEITEC) Nano Research Infrastructure. E.P.d.A. acknowledges financial support from the European Union's Horizon 2020 research and innovation programme under the Marie Skłodowska-Curie grant agreement No. 101105733. We thank M. Kovařík for additional AFM measurements, and L. Berners for help with the spherical harmonics band profile analyses.

# Associated Content

This manuscript has been previously submitted as a preprint as: Zhang, T; Holzer, J; Vystavěl, T; Kolíbal, M; de Araujo, E. P.; Stephens, C; Britton, T. B. Characterization of $WSe_2$ films using reflection Kikuchi diffraction in the scanning electron microscope and multivariate statistical analyses (2025). arXiv:2506.22575, https://arxiv.org/abs/2506.22575. (Accessed October 16, 2025)

**Supporting Information:**

Additional details and graphics of the experimental set-ups, parameters and data processing routines, mathematical formulation of the MSA approach, and supplementary results from the MSA analysis (DOC).

## Graphics for Table of Contents

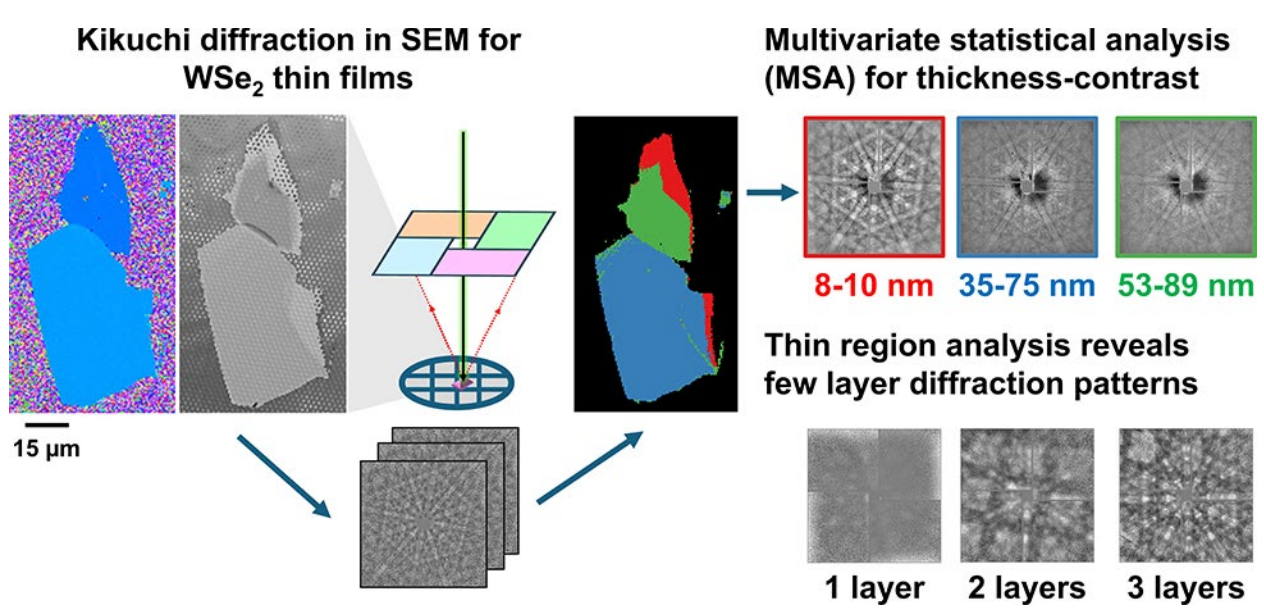

Supplementary Information

# Characterization of thin films using reflection Kikuchi diffraction in the scanning electron microscope and multivariate statistical analyses

Tianbi Zhang[1], Jakub Holzer[2], Tomáš Vystavěl[2], Miroslav Kolíbal[3,4], Estacio Paiva de Araujo[3], Chris Stephens[5], T. Ben Britton[1]

1. Department of Materials Engineering, University of British Columbia, 309-6350 Stores Road, Vancouver BC, V6T 1Z4 Canada
2. Thermo Fisher Scientific, Vlastimila Pecha 1282/12, 627 00, Brno, Czech Republic
3. Central European Institute of Technology, Brno University of Technology, Purkynova 123, CZ-61200 Brno, Czech Republic
4. Institute of Physical Engineering, Brno University of Technology, Technická 2, 616 69 Brno, Czech Republic
5. Thermo Fisher Scientific, Materials Science, East Grinstead, United Kingdom

**RKD and EBSD experiment parameters.**

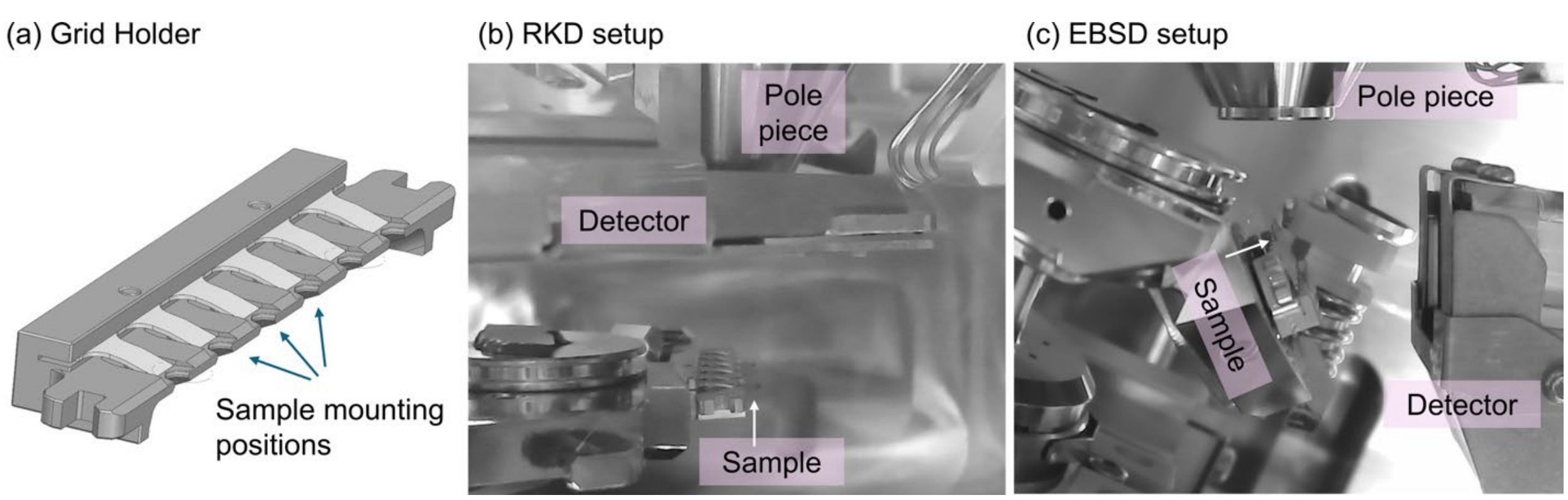

Figure S1. (a) CAD rendering of the TEM grid holder for sample mounting. (b) and (c) shows in-chamber photos of the setups of RKD and EBSD, respectively, with key components highlighted and labelled.

Experimental conditions and parameters of RKD and EBSD maps are listed in Table S1.

Table S1. Parameters used in the RKD and EBSD mapping experiments.

| Primary electron beam energy (keV) | Energy threshold (keV) | Beam current (nA) | Exposure time (ms) | Incident Electron Dose per Pattern ($10^5$ electrons) | Map step size (nm) | Map size (points) |
|---|---|---|---|---|---|---|
| Multilayer mapping - RKD | | | | | | |
| 7 | 5 | 12.8 | 20 | 16 | 566 | 116x209 |
| 10 | 6 | 6.4 | 20 | 8 | 566 | 100x192 |
| 15 | 11 | 6.4 | 20 | 8 | 566 | 116x209 |
| 20 | 16 | 6.4 | 30 | 12 | 566 | 110x197 |
| Multilayer mapping - EBSD | | | | | | |
| 10 | 5 | 6.4 | 10 | 4 | 500 | 113x180 |
| Few-layer mapping – RKD | | | | | | |
| 8 (monolayer) | 4 | 6.4 | 50 | 20 | 100 | 78x73 |
| 8 (bilayer) | 4 | 3.2 | 50 | 10 | 50 | 66x64 |
| 8 (trilayer) | 4 | 3.2 | 50 | 10 | 50 | 91x148 |

**Full MSA data on RKD and EBSD.** Figure S2 shows the score maps, characteristic patterns and assignment map of all 13 components identified.

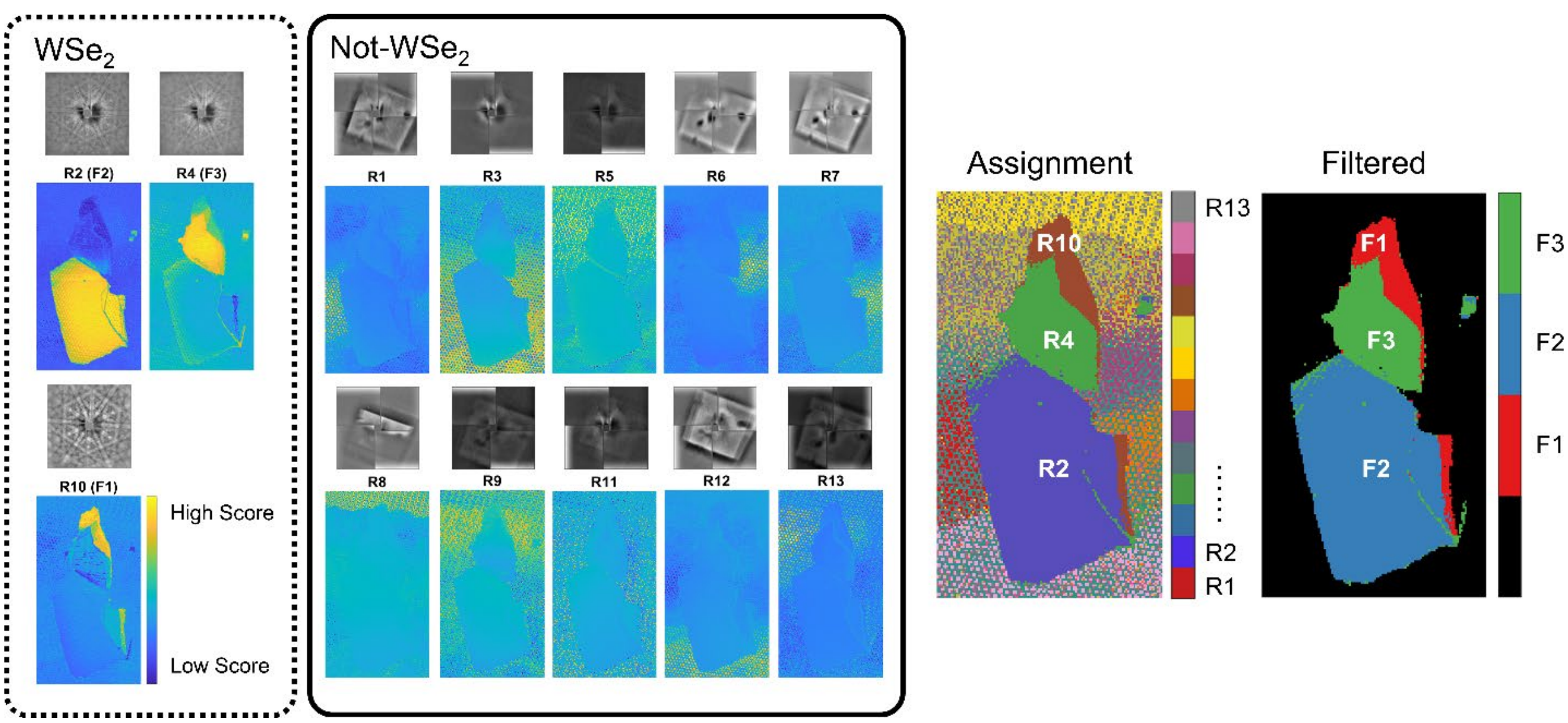

Figure S2. Score maps and characteristic patterns of the 13 components identified by MSA based on RKD data collected at 7 keV from the multilayer film.

For these ‘Not-$WSe_2$’ “characteristic patterns”, many show the shadow of the TEM grid and/or the substrate that is used to hold the sample, and the shadow is formed as a projection from electrons striking lower down in the microscope, backscattering and then the shadow pattern is collected by the RKD detector. Within the score map, additionally the holes and substrate of the holey grid can be seen as well.

**MSA on EBSD dataset.** Figure S3 shows the score maps, characteristic patterns and assignment map of all 13 components identified.

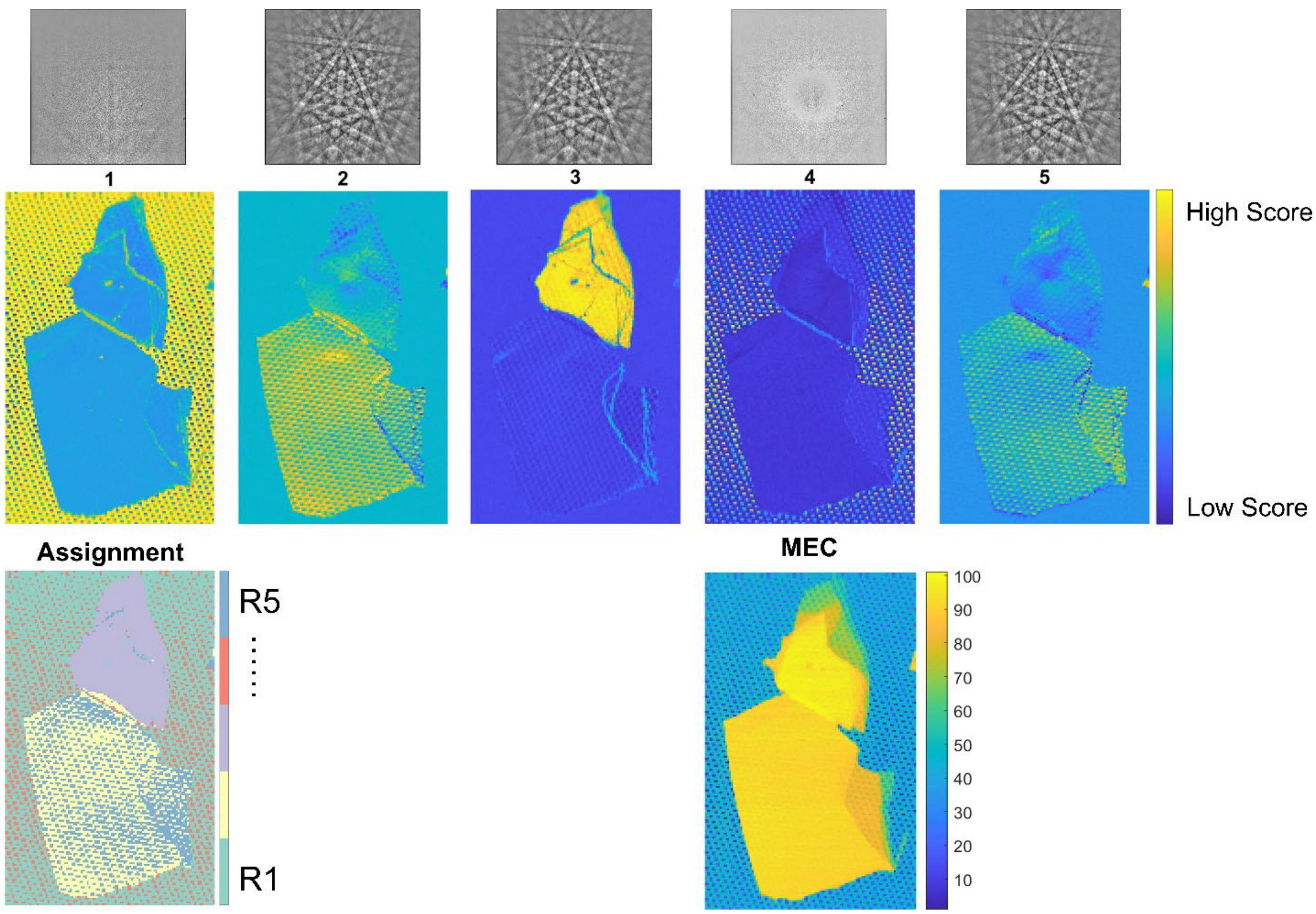

Figure S3. Score maps and characteristic patterns of the 5 components identified by MSA based on EBSD data collected at 10 keV from the multilayer film.

**MSA on Monolayer sample.** As an additional approach to enhance our understanding, we have also applied the MSA approach to a subregion of Figure S4, where part of the monolayer film was in contact with the amorphous TEM grid and the rest above a hole of the grid, where optical micrograph already reveals contrast differences. Here, using a variance tolerance of 0.025%, it is possible to separate regions which correspond to film supported by the grid and film over a hole on the grid.

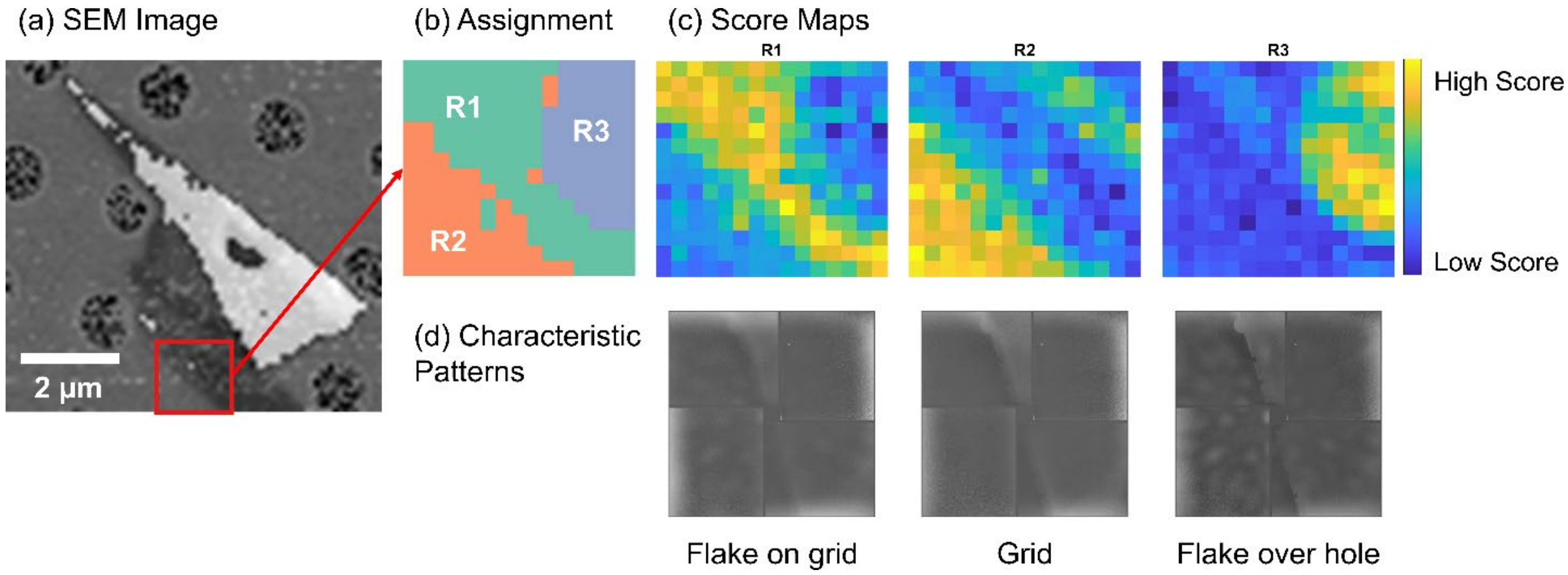

Figure S4. Score maps and characteristic patterns of the 3 components identified by MSA based on RKD data collected at 8 keV from the monolayer film sample.

The difference between the patterns from the flake in contact and not in contact with the gris echoes the similar observation of Figure 6 in the main manuscript.

**Mathematical formulation of the MSA approach.** The multivariate statistical analysis approach is based first on principal component analysis (PCA). In PCA, an orthogonal set of linearly uncorrelated variables ("principal components") are calculated based on singular value decomposition of the input data matrix [1]. The principal components are ordered so that the first principal component has the largest inter-class variance in the data, and so on. For a Kikuchi diffraction dataset with $s \times t$ data points, with each point associated with a $m \times n$ pixel diffraction pattern, a total of $m \times n$ principal components will be obtained. We can select the strongest *p* components and classify the dataset based on the strongest components, or "characteristic patterns", and then the strengths of each component can be calculated by:

$$\boldsymbol{D}_{[\mathrm{m\times n,s\times t}]} = \boldsymbol{C}_{[\mathrm{m\times n,p}]}\,\boldsymbol{S}^{\mathrm{T}}_{[\mathrm{p,s\times t}]} \tag{1}$$

Where ***D*** is the data matrix (with the diffraction patterns vectorized), ***C*** is the set of principle components, and ***S*** contains the spatial map of the strengths of basis pattern signal, or the "score map".

Wilkinson et al. [2] note that for an EBSD dataset, the principal components are mixtures of experimental patterns, and it is not guaranteed that a data point will be dominated by a characteristic pattern, and data clustering or classification tend to be ineffective or unmeaningful [2,3]. Wilkinson suggested use of PCA followed by the VARIMAX rotation, $\boldsymbol{R}$, of the basis vectors solution [4], such that:

$$\boldsymbol{D}_{[\mathrm{m}\times\mathrm{n},\mathrm{s}\times\mathrm{t}]} = \boldsymbol{C}_{[\mathrm{m}\times\mathrm{n},\mathrm{p}]}\boldsymbol{R}_{[p,p]}\boldsymbol{R}^{T}_{[p,p]}\ \boldsymbol{S}^{\mathrm{T}}_{[\mathrm{p},\mathrm{s}\times\mathrm{t}]} \tag{2}$$

**Effect of pattern processing on MSA.** A particular advantage of MSA is to amplify the signal-to-noise ratio of the raw data [2,5] for more effective non-local data clustering. Therefore, pre-processing of the data, such as background correction and binning, should enhance the MSA results. This is confirmed in Figure S5 by the same MSA routine on (1) full-sized, background corrected patterns and (2) full-sized raw patterns from the 10 keV dataset. The same variance tolerance of 0.01% was used. MSA on full-sized, background corrected pattern also identified 13 components with very similar result of clustering to that using binned patterns. On the other hand, using raw patterns, only 2 components were identified, which roughly correspond to the thicker regions on the flake and the rest, while thin regions on the flake are wrongly classified along with non-diffracting features.

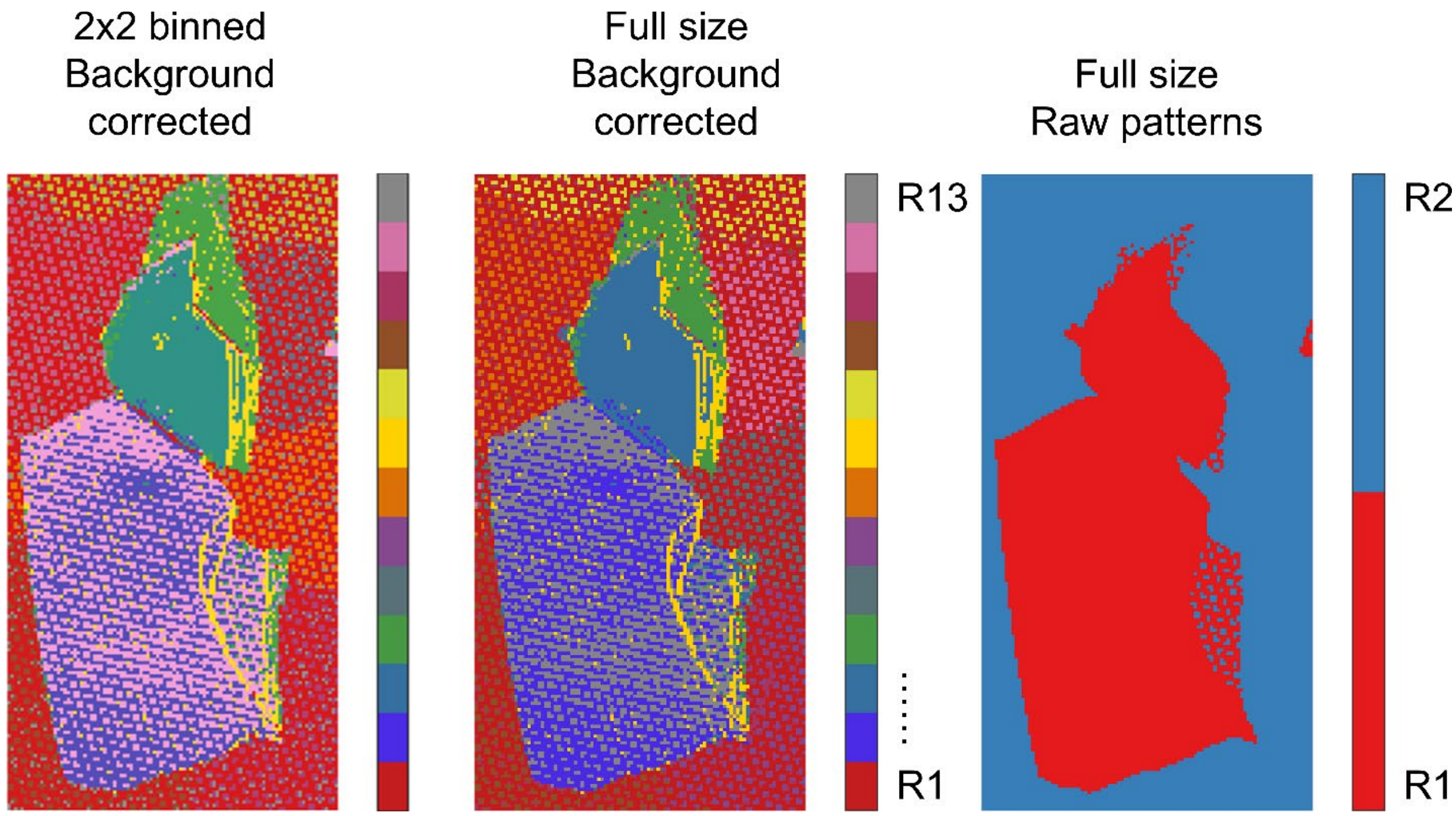

Figure S5. Assignment maps from MSA on 10 keV RKD data using background corrected patterns with 2x2 binning, background corrected patterns with no binning, and raw patterns. $WSe_2$ layers support is 200 nm holey SiN membrane.